\documentclass [12pt,a4paper]{article}
\usepackage{amssymb, theorem}
\newtheorem{lemma}{Lemma}
\newtheorem{theorem}{Theorem}
\theorembodyfont{\rm}
\newtheorem{example}{Example}
\def\qed{\nobreak\hfill $\square$}
\newcommand{\<}{\langle}
\renewcommand{\>}{\rangle}

\def\proof{\noindent{\it Proof.} }

\def\bbbr{{\mathbb R}}
\def\bbbc{{\mathbb C}}
\def\eps{\varepsilon}

\def\Tr{\mathrm{Tr}\,}
\def\Var{\mathrm{Var}\,}

\def\im{\mathrm{i}}

\def\Det{\mbox{det}\,}

\def\ep{\varepsilon}

\def\Mn{M_n(\bbbc)}

\def\iA{{\mathcal A}}
\def\iB{{\mathcal B}}
\def\iH{{\mathcal H}}

\def\ot{\otimes}
\def\osum{\oplus}

\begin{document}
\rightline{Rep. Math. Phys.}
\vskip 2cm
\centerline{\LARGE {\bf Efficient quantum tomography}}
\medskip
\centerline{\LARGE {\bf  needs complementary and  symmetric}}
\medskip
\centerline{\LARGE {\bf  measurements}}
\bigskip
\bigskip
\centerline{D\'enes Petz  
\footnote{E-mail: petz@math.bme.hu.} and
L\'aszl\'o Ruppert 
\footnote{E-mail: ruppertl@math.bme.hu.}}
\bigskip
\begin{center}
Alfr\'ed R\'enyi Institute of Mathematics\\
H-1053 Budapest, Re\'altanoda u. 13-15, Hungary  
\end{center}
\begin{center}
Department for Mathematical Analysis, \\
Budapest University of Technology and Economics\\
H-1521 Budapest XI., Hungary
\end{center}

\bigskip
\begin{abstract}
In this study the determinant of the average quadratic error matrix is
used as the measure of state estimation efficiency. Minimizing this quantity 
gives us the optimal
measurements in different scenarios. We present applications when von Neumann 
measurements or a single POVM are used,
when there is no known information or a part of the parameters of the
state is given. Under some restrictions the optimality is found for $n$-level
systems. The optimal measurements have some complementary relation
to each other and to the available data, moreover, symmetric informationally
complete systems appear, containing a new, conditional version.

\bigskip
\medskip\noindent
{\bf Key words and phrases:} experiment design, state estimation,
complementarity, measurement, quadratic error, qubit, symmetric informationally
complete POVM.

\end{abstract}
\bigskip 

\newpage
\section{Introduction}

State estimation or tomography is a fundamental problem in the field of quantum
information theory and it can be considered as one of the
foundational issues of quantum mechanics \cite{tomography,PDbook}.
The topic of quantum tomography consists of methods which reconstruct the state 
of the system under investigation using repeated measurements of a set of observables. 
The problem may be traced back to the seventies \cite{Helstrom},
the interest in a thorough mathematical analysis of the
quantum state estimation procedures has been flourishing recently
\cite{Bagan, hayashi, design, REK}. 

In statistics the accuracy of the estimation can be quantified by
the quadratic error matrix. The matrices are typically not
comparable by the positive semi-definiteness, hence if different
estimation schemes are compared, the determinant of mean quadratic
error matrix can be used instead. This approach can be found in references 
\cite{PeHa:2007, PeHa:2008}, their main result was
that the complementary von Neumann measurements are optimal. A more
general context appears in \cite{BP}, which is much closer to our
approach. The concept of complementarity was extended to
quantum subsystems in \cite{PDcomp} and the case of two qubits has
a detailed analysis in \cite{OPSz, m4}. When all parameters of the
density matrix are obtained from a single measurement, then a
symmetric informationally complete POVM appears \cite{Renes}.
A similar result was obtained earlier by Wootters and
Fields \cite{WoFi} for von Neumann measurements and by Scott \cite{Scott} for POVMs,
but optimality had a different formulation in both cases.

Some a priori information about the state can be given in various ways, for example 
one can minimize the relative entropy to an a priori given state, resulting in a 
bias toward that state \cite{oliv}. The most popular subject in this field is 
state discrimination: when we know that the system is one of several given states 
and we should figure out which one \cite{dar05}. Beside knowing the possible states 
we can have an a priori probability distribution on the true state, too. This idea 
was used in \cite{demko} to obtain the optimal phase estimation. The given states 
do not construct a discrete set, instead, they are searching among all pure states. 
Another continuous subset is used in \cite{konrad}, where we know that the state 
to estimate is a convex combination of two given states. In our setup we know that 
the state is on a given subset of the whole state space (this can be the whole 
state space itself) and assume an arbitrary unitarily invariant a priori probability 
distribution to be known. 

We use the determinant of the average
quadratic error matrix as a figure of merit in state estimation setup; the average is
taken over the unitarily invariant states. This quantity is relies on purely 
classical statistics, therefore it is
much easier to compute than, e.g., quantum Fisher information, but still reproduces 
the well-known quantum phenomena and even more.
So it gives us a useful tool to find the optimal measurement setup for different quantum 
tomography problems. We examine the assumption that
the unknown state is known on some complementary subalgebras,
which practically means that one should restrict the are of integration.
For our calculations we use a parametrization for both the density operators and the 
observables (we expand them
in a fixed basis that is orthonormal with respect to the Hilbert-Schmidt inner product), 
but we can see that the results are independent of this parametrization. To gain analytic
 results, some sort of symmetry is recommended  but by using numerical methods
we could overcome that and solve difficult problems efficiently.

Section 2 is a very short introduction, quantum states, measurements, state estimation, 
complementarity and symmetric
measurements are overviewed. In Section 3, we give the motivation for using the determinant 
of the average covariance matrix by guiding through a simple example, and then give some 
useful remarks about this quantity. Then the optimization problem is solved for arbitrary 
dimensions under some restrictions. In Section 4 the complementarity of optimal von 
Neumann measurements is obtained. In Section 5 a single POVM is used for the measurement, 
then a symmetric informationally complete POVM appears. Finally, we investigate the 
asymmetrical case for POVMs too (Sec. 6) and draw conclusions (Sec. 7).

\section{Basic concepts}

In this section, we give a brief summary of the basic concepts appearing in the 
present paper (for further details, see \cite{tomography,PDbook}).

$\rho \in M_n(\bbbc)$ is an n-dimensional quantum state if
\begin{equation}\label{state}
\rho \ge 0 \quad \textrm{ and } \quad \Tr \rho=1.
\end{equation}

We will use a basis of self-adjoint matrices to obtain a
parametrization of these density matrices. In the qubit case, we use the
Pauli matrices as a basis, i.e., the Bloch parametrization:
\begin{equation}\label{E:tetak}
\rho=\frac{1}{2} \left( I + \sum_{i=1}^3 \theta _i \sigma_i\right)=
\frac{1}{2} \left( I + \theta \cdot \sigma \right)=
\frac{1}{2} \left[ \begin{array}{cc}
1+\theta _3& \theta _1-\im \theta _2\\ \theta _1+\im \theta _2 &1-
\theta _3\end{array} \right].
\end{equation}

In this case the whole state space of quantum states can be described using the Bloch
vector: $\theta=(\theta_1,\theta_2,\theta_3)$ and the conditions in (\ref{state}) are 
simply converted into $\theta _1^2+\theta _2^2+\theta _3^2\le 1$.

For the n-dimensional case we use the following parametrization:
$$
\rho=\frac1n (I + \theta \cdot \sigma)
$$
where $\theta \in \bbbr^{n^2-1}$ is the generalized Bloch vector, $\theta\cdot \sigma=\sum_j 
\theta_j \sigma_j$ and $\{\sigma_j: 1 \le j \le n^2-1\}$ are generalized Pauli matrices, 
which form an orthonormal basis on the self-adjoint traceless matrices: $\sigma_i=\sigma_i^*$, 
$\Tr \sigma_i=0$ and $\Tr \sigma_i \sigma_j=\delta_{i,j}$. An example for the  3-dimensional 
case are the multiples of Gell-Mann matrices.

A positive operator-valued measure (POVM) can in this framework be described as a set of 
matrices: 
$$
E=\{E_1,E_2,\dots E_k\}\quad \mbox{with}\quad \sum\limits_{i=1}^k E_i=I, E_i>0. 
$$
Von Neumann measurements can be described as a POVM with two components and projective elements: 
$\{P_1, I-P_1\}$.

Having a quantum state $\rho$ and a POVM $E$, the probability of obtaining an outcome related 
to the  POVM element $E_i$ is:
\begin{equation}\label{measureprob}
p_i=\Tr(\rho E_i).
\end{equation}
If we have $m$ samples of the state $\rho$ and we measure $E$ on each copy independently, 
then let $\nu_i$ be the relative frequency of measuring the $i$th outcome. If $m$ is a 
large number, then $\nu_i$ will be close to $p_i$, and if we know the POVM (because we 
set it), then we can make an estimation on $\rho$ from (\ref{measureprob}) by using
$\nu_i$ instead of $p_i$. A simple state estimation problem is described in detail in 
Section \ref{sec:DACM}.

The heuristic concept of complementarity was born together with quantum theory.
A mathematical definition is due to Accardi \cite{Acc} and Kraus \cite{Kr}. Let
$\iH$ be an $n$-dimensional Hilbert space. Let the observables $A$ and $B$ have
eigenvectors $e_1,e_2,\dots, e_n$ and $f_1,f_2,\dots, f_n$ which are orthonormal
bases, they are complementary if
\begin{equation}\label{E:c}
|\< e_i, f_j\>|^2=\frac{1}{n}\qquad (1 \le i,j \le n).
\end{equation}
If this condition holds then the two bases are also called mutually unbiased.
An overview about complementarity can be found in \cite{jmp} and details are in \cite{PDcomp, m4}.

Complementarity can be generalized to the case of POVMs. The POVMs $\{E_1,E_2,$ $\dots, E_k\}$
and $\{F_1,F_2,$ $\dots, F_m\}$ are complementary if
\begin{equation}\label{E:c2}
\Tr E_i F_j=\frac{1}{n}\Tr E_i \, \Tr F_j \qquad (1 \le i \le k, \quad 1 \le j \le m).
\end{equation}
This is equivalent to the orthogonality of the traceless parts:
$$
E_i - \frac{\Tr E_i}{n} I \quad \perp \quad F_j - \frac{\Tr F_j}{n} I,
$$
so we will use the expression quasi-orthogonal, when we use this property of 
complementary operators.

The concept can be extended to subalgebras $\iA_1,\iA_2 \subset \Mn$. $\iA_1$ and $\iA_2$
are complementary if  $A_1 \in \iA_1$ and $A_2 \in \iA_2$ are quasi-orthogonal (that is, 
the traceless part of matrices are orthogonal).

An orthonormal basis corresponds to the subalgebra of operators which are diagonal
in this basis. Such a subalgebra is maximal Abelian, it will be called M-subalgebra.

A symmetric informationally complete POVM $\{E_i\,:\, 1 \le i \le k\}$ of an $n$-level
system is described by a set of projections $P_i$ ($1 \le i\le k$) such that
\begin{equation}\label{E:SIC}
\sum_{i=1}^k P_i=\lambda I \quad \mbox{and} \quad \Tr P_i P_j=\mu \quad (i\ne j),
\end{equation}
where $k=n^2$, $\lambda=n$, $\mu=1/(n+1)$ and all projections are rank-one, see 
\cite{Renes, Ruskai}. The existence of a symmetric informationally complete POVM (SIC POVM) 
is not known for every dimension
$n$.

In this paper $k < n^2$ case will also appear, when some parameters are already known, 
we need to estimate fewer parameters, i.e., fewer elements in POVM are needed to make 
a full reconstruction. If they fulfill condition (\ref{E:SIC}) with some constants and 
the $P_i$-s are complementary to the known subalgebra, then we call them conditional 
symmetric informationally complete POVM toward the known $n^2-k$ parameters, because 
they are informationally complete only if we use the condition that we want to discriminate 
the states only in the unknown directions. They include the case of SIC-POVMs as a special 
case when no parameters are known. The existence of such a POVM can be a fundamental 
question in different quantum tomography problems.

\section{The determinant of the average covariance matrix}\label{sec:DACM}

Suppose that we have a qubit and some of
the three parameters $\theta _1$, $\theta _2$ and $\theta _3$ are known and
the unknown parameters will be estimated. For example, let us assume that $\theta_1$ and  
$\theta_2$ are known and we want to
estimate $\theta_3$. The assumption means that the reduced state is known
on the M-subalgebras generated by $\sigma_1$ and $\sigma_2$, respectively.
They are complementary subalgebras and the M-subalgebra generated by $\sigma_3$
is complementary to both. A projection of a von Neumann measurement
$$
E=\frac{1}{2} \left( I + \lambda \cdot \sigma\right)
$$
is used to estimate $\theta _3$ from the result of several measurements, where
$\lambda_1^2+\lambda_2^2+\lambda_3^2 = 1$. The expected value
is
$$
\Tr \rho E=\frac{1}{2}\left(1+\sum_{i=1}^3 \theta _i \Tr \sigma_i E \right)
=\frac{1}{2} \left( 1 + \<\theta , \lambda\>\right)
$$
and we have
$$
\theta_3\Tr \sigma_3 E =2\Tr \rho E- \left(1+\sum_{i=1}^2 \theta_i
\Tr \sigma_i E \right)\,.
$$
Let us denote by $\nu$ the random outcome of the measurement of $E$, the expected value
is $p=\Tr \rho E$ and it is Bernoulli distributed. The natural unbiased estimate of
$\theta _3$ is
\begin{equation}\label{E:esti}
\hat{ \theta} _3 =\frac{1}{\Tr \sigma_3 E}\left(
2\nu - 1-\sum_{i=1}^2 \theta _i \Tr \sigma_i E \right).
\end{equation}
The variance is
\begin{equation}\label{E:var}
\Var(\hat{ \theta} _3)=\frac{4p(1-p)}{\lambda_3^2}=\frac{1-\<\lambda, \theta \>^2}{\lambda_3^2}\,.
\end{equation}

We want to argue that the optimal estimate corresponds to the projection $\lambda_1=
\lambda_2=0, \quad \lambda_3=\pm 1$, i.e. measuring in the unknown direction. This estimator
has the variance $1-\theta _3^2$. However, the inequality
\begin{equation}\label{E:int2}
\frac{1-\<\lambda, \theta \>^2}{\lambda_3^2}\ge 1-\theta _3^2
\end{equation}
is not true in general, for example $\lambda=\theta $ is possible
for pure states and then the left-hand side is zero. So if we want to minimize 
$\Var(\hat{ \theta}_3)$, it is not true that $\frac{I+\sigma_3}{2}$ is the best observable.

Since $\theta_1, \nobreakspace \theta_2$ are given, the only unitarily
invariant vector to $(\theta_1,\theta_2,\theta_3)$ is
$(\theta_1,\theta_2,-\theta_3)$, if we take the average of the left hand side of (\ref{E:int2}) 
on these states we obtain
$$
\left<\Var(\hat{ \theta} _3)\right>=\frac12\cdot \frac{1-(\lambda_1 \theta_1+\lambda_2 \theta_2
+\lambda_3 \theta_3)^2}{\lambda_3^2}+\frac12\cdot \frac{1-(\lambda_1 \theta_1+\lambda_2 \theta_2
-\lambda_3 \theta_3)^2}{\lambda_3^2}.
$$
We can solve the minimization problem
$$
\left<\Var(\hat{ \theta} _3)\right>\rightarrow \textrm{min}, \quad \lambda_1^2+\lambda_2^2
+\lambda_3^2 = 1
$$
using the Schwarz inequality, and get that in the optimal case $\lambda_1=\lambda_2=0, 
~\lambda_3=\pm 1$. Thus, the measurement of $\frac{I+\sigma_3}{2}$ is optimal indeed, 
yet only in the average sense.

Note that in (\ref{E:esti}) $\nu$ has Bernoulli distribution, with $p(1-p)$ variance.
If $m$ measurements are performed, we will have $\nu=(\nu_1+\nu_2+\dots+\nu_m)/m$, where 
$\nu_i$ is the outcome of the $i$-th
measurement. Then the variance will be $\frac{m p(1-p)}{m^2}=\frac{1}{m} p(1-p)$, so we 
can conclude that if we have multiple measurements the variance changes only by a constant 
factor, hence the minimization problem remains unaffected.

In the following sections, we will always use the same kind of
naturally derived estimators; one can prove that those are
unbiased and efficient.  
We should also always bear in mind that if we have only a few measurements, we can 
easily get a Bloch-vector as a result
of the estimation, which does not satisfy the positivity condition in (\ref{state}). But in
practical cases, we have many measurements, hence the law of large
numbers ensures us that the estimator will give a state near
the original one, and the probability of getting a physically
impossible state (a density matrix with at least one negative eigenvalue) will converge 
to zero exponentially, due to the
large deviation theory.
That is why we do not have to deal with any
physical restrictions, even if we calculate the variance only for one
measurement in each case.

If we have multiple parameters, we can still get a multidimensional estimator 
from the measurement outcomes, but in that case, a covariance matrix will take the role 
of the variance. The average covariance matrix can be calculated 
by taking the piecewise average of
the elements, but since the matrices are not comparable in general, we need an 
$M_n(\bbbc)\rightarrow \bbbr$ function which gives us a number
as the objective function to minimize. After examining some candidates we concluded that 
the best choice is to take the determinant of the matrix and that this needs to be done after the 
averaging process, so we have the following optimization problem:
$$
\det\left<Cov\left( \underline{\hat{\theta}}\right)\right>\rightarrow \textrm{min},
$$
where $\underline{\hat{\theta}}$ is the estimator of the unknown parameters.
This way we obtain an easily calculable quantity, and by solving the minimization problem, 
the optimal measurements.

If we want to take the average of this covariance matrix on the whole state space,
we have to give a unitarily invariant measure on the state space. On the other hand, 
if we integrate on the
unitarily invariant states of an initial state with respect to the normalized Haar measure 
and the minimum is independent of the initial state, we get optimal measurements
independently from the (unitarily invariant) probability measure. Thus,
 we can obtain a stronger result than we would by using a particular measure.

Finally, let us note that using a POVM instead of von Neumann measurements is not a crucial 
difference. If there are $n$ unknown parameters, we need to have
$n$ independent positive operators in the POVM. We only have the restriction $\sum E_i=I$, 
so $n+1$ operators are sufficient and we can construct the estimator from the outcomes 
related to the first $n$ components. The only difference is that in this case, the $n$ 
outcomes would not be independent, off-diagonal elements appear in the covariance matrix, too.

\section{The optimal von Neumann measurement}

The $n$-level quantum system is described by the algebra $M_n(\bbbc)$.
Consider the following decomposition
$$
M_n(\bbbc)=\bbbc I \osum \iA \osum \iB,
$$
where $\iA$ and $\iB$ are linear subspaces and orthogonality is defined
with respect to the Hilbert-Schmidt inner product $\<A,B\>=\Tr A^*B$.

A state has the density matrix
\begin{equation}\label{E:deco}
\rho=I/n + \rho_\iA + \rho_\iB.
\end{equation}
We assume that the component $\rho_\iB$ should be estimated. Let
the dimension of $\iB$ be $k$. The positive contractions $E^1,
\dots,E^k$ are used for independent measurements (on several
identical copies of the $n$-level system): A measurement
corresponds to the POVM $\{E^i, I-E^i\}$. These operators have the
expansion
\begin{equation}\label{decomp}
E^i=e_iI + E^i_\iA+ E^i_\iB \qquad (1 \le i \le k).
\end{equation}
The expectations are
$$
p_i:=\Tr \rho E^i=e_i + \Tr \rho_\iA E^i_\iA +\Tr \rho_\iB E^i_\iB
\qquad (1 \le i \le k).
$$

We fix an orthonormal basis $F_1,\dots,F_k$ in $\iB$. The unknown component
has the expansion
$$
\rho_\iB=\theta_1 F_1+\dots +\theta_k F_k,
$$
where $\theta=(\theta_1,\dots, \theta_k)$ are the parameters to be estimated.
Similarly,
$$
E^i_\iB=e_{i1} F_1+ e_{i2} F_2+\dots +e_{ik} F_k
$$
is an orthogonal expansion.

The estimates $\hat \theta_i$ are solutions of the equations
$$
\eps_i=e_i + \Tr \rho_\iA E^i_\iA + \sum_{j=1}^k e_{ij} \hat \theta_{j},
$$
where $\eps_i$ is the random result of the $i$th measurement, $1 \le i \le k$. In
another form
$$
\left[\matrix{ \eps_1 \cr \vdots \cr \eps_k }\right]=
\left[\matrix{ e_1 \cr \vdots \cr e_k}\right]+
\left[\matrix{  \Tr  \rho_\iA E^1_\iA \cr \vdots \cr
\Tr \rho_\iA E^k_\iA}\right]+
\left[\matrix{  e_{11} & \cdots & e_{1k} \cr
\vdots & \ddots & \vdots \cr e_{k1} &\cdots & e_{kk}}\right]
\hat \theta^t
$$
or in a different notation
$$
\left[\matrix{ \eps_1 \cr \vdots \cr \eps_k }\right]=
\left[\matrix{ e_1+ \Tr  \rho_\iA E^1_\iA\cr \vdots \cr
e_k+ \Tr  \rho_\iA E^k_\iA  }\right]+T \hat \theta^t\, ,
$$
where $T$ is the $k \times k$ matrix from the previous formula. Therefore,
$$
\hat \theta^t=
T^{-1}
\left(\left[\matrix{ \eps_1 \cr \vdots \cr \eps_k }\right]-
\left[\matrix{ e_1+ \Tr  \rho_\iA E^1_\iA\cr \vdots \cr
e_k+ \Tr  \rho_\iA E^k_\iA  }\right]\right)\,.
$$
We have
$$
\hat \theta^t- \theta^t =T^{-1}
\left(\left[\matrix{ \eps_1 \cr \vdots \cr \eps_k }\right]-
\left[\matrix{ p_1 \cr \vdots \cr p_k }\right]\right).
$$
The quadratic error matrix is the expected value of
$$
T^{-1}\left[\matrix{ \eps_1-p_1 \cr \vdots \cr \eps_k-p_k }\right]
\left[\matrix{ \eps_1-p_1, &  \dots & , \eps_k-p_k }\right]
(T^{-1})^t=
T^{-1}\Big[(\eps_i-p_i)(\eps_j-p_j)\Big]_{i,j=1}^k(T^{-1})^t.
$$
Due to the independence of the measurements the expected value of
$$
\Big[(\eps_i-p_i)(\eps_j-p_j)\Big]_{i,j=1}^k
$$
will be diagonal.

The random variables $\eps_i$ have Bernoulli distribution and their variance is
$$
(1-\Tr \rho E^i)\Tr \rho E^i.
$$
We want to take the average:
$$
\int (1-\Tr \rho E^i)\Tr \rho E^i\,d\mu(\rho),
$$
where we integrate on the unitarily invariant states, and $\mu$ is the corresponding 
normalized Haar measure.
For the sake of simplicity, assume that the operators $E^i$ have the same spectrum.
Then the integral is constant: it does not depend on the actual $E_i$, so the average 
of the quadratic error matrix is

$$
c T^{-1}(T^{-1})^t.
$$
The determinant is minimal if the determinant of the matrix $T$ is
maximal. Geometrically, the determinant is the volume of the
parallelepiped determined by the row vectors. To maximize the
determinant, the row vectors should be long. This implies that $E^i_\iA=0$ in (\ref{decomp}), 
since otherwise we could project
$E^i$ on to $\bbbc I \osum \iB$, and still have the same elements in $T$. So we have
$$
E^i=e + e_{i1} F_1+ \dots +e_{ik} F_k \qquad (1 \le i \le k),
$$
and then the determinant of the matrix
$$
\left[\matrix{ 1/n & 0 & \cdots & 0 \cr
e & e_{11} & \cdots & e_{1k} \cr
\vdots & \vdots & \ddots & \vdots \cr
e & e_{k1} & \cdots & e_{kk}}\right]
$$
is $\Det T$. The angle of the first row and any other row is
fixed. To have a large determinant the rows of $T$ should be
orthogonal. In this case, the operators $E^1,\dots,E^k$ are
quasi-orthogonal.

\begin{theorem}\label{nlevel}
If the positive contractions $E^1,\dots,E^k$ have the same spectrum, then the
determinant of the average of the quadratic error matrix is minimal if the
operators $E^1,\dots,E^k$ are complementary to each other and to $\iA$.
\end{theorem}

\begin{example}
If two qubits are given and the reduced states of both qubits are known, then the
ideal state estimation is connected to the observables
$$
\sigma_{11},\sigma_{22},\sigma_{33}, \sigma_{12},\sigma_{23},\sigma_{31},
\sigma_{13},\sigma_{21},\sigma_{32},
$$
where $\sigma_{ij}=\sigma_i \ot \sigma_j$.  \qed
\end{example}

\section{The optimal POVM}

Once again we are in the n-dimensional case, but in this section we will use POVMs for 
state estimation.
Assume that $\iA=\{0\}$, which means that all parameters are unknown. Then
$$
E_i=e_iI + E^i_\iB= e_i(I + f_i\cdot \sigma) \qquad (1 \le i \le n^2),
$$
where $e_i \in \bbbr$, $f_i \in \bbbr^{n^2-1}$, $f_i\cdot \sigma=\sum_j f_{ij}\sigma_j$ and
$\{\sigma_j: 1 \le j \le n^2-1\}$ are generalized Pauli matrices, $\Tr \sigma_i=0$,
$\Tr \sigma_i^2=1$,  $\Tr \sigma_i \sigma_j=0$. The positivity condition of $E_i$ is
not known, but we have a necessary condition:

\begin{lemma}\label{L:poz}
If the $n\times n$ matrix $I+ g\cdot \sigma$ is positive, then $\sum_k g_k^2 \le n^2 -n$.
If $\sum_k g_k^2 = n^2 -n$, then $I+ g\cdot \sigma= nP$ with a projection $P$ of rank 1.
\end{lemma}

\proof
$A=g\cdot \sigma$ is self-adjoint, $\Tr A= 0$. Let $\lambda_1,\lambda_2, \dots, \lambda_n$
be the eigenvalues of $A$. Then
$$
\sum_{k=1}^{n^2-1} g_k^2=\Tr A^2=\sum_{t=1}^n \lambda_t^2.
$$
Since $\sum_{t=1}^n \lambda_t=0$ from $\Tr A=0$ and $\lambda_t \ge -1$ from $I+A \ge 0$, we
have the upper bound. Namely, $\sum_{t=1}^n \lambda_t^2$ is maximal, if $\lambda_1,\lambda_2,
\dots, \lambda_n$ is a permutation of the numbers $-1,-1, \dots, -1, n-1$.  In this
case $I+A$ has eigenvalues $0,0, \dots, 0, n$, so it is a multiple of a projection.  \qed

The computations are similar to the von Neumann case. The probabilities of different outcomes are
$$
\left[\matrix{p_1 \cr\vdots \cr p_d }\right]=
\left[\matrix{e_1 \cr \vdots \cr e_d }\right]+T
\left[\matrix{ \theta_1 \cr \vdots  \cr \theta_d }\right]
$$
where $d=n^2-1$ and $T=(e_1, e_2,\dots, e_{d})^T (f_1, f_2,\dots, f_{d})$ dyadic matrix.

If $\nu_1,\nu_2,\dots \nu_d$ are the relative frequencies of the outcomes (of different
measurements on identical copies), then the estimator is
\begin{equation}\label{E:sol2}
\left[\matrix{ \hat\theta_1 \cr \vdots  \cr \hat\theta_d }\right]=T^{-1}
\left[\matrix{\nu_1-e_1 \cr \vdots \cr \nu_d-e_d}\right].
\end{equation}

The mean quadratic error matrix is
$$
V(\theta)=T^{-1}\,W\,(T^{-1})^*
$$
with $W$ denoting the covariance matrix of the random variables $\nu_1,\nu_2,\dots \nu_d$. 
These are multinomially distributed, so
$$
W=\left[\matrix{p_1(1-p_1) & - p_1 p_2 & \dots &  - p_1 p_d\cr
 - p_1 p_2 & p_2(1-p_2) &\dots &  - p_2 p_d\cr
\vdots& \vdots &&\vdots \cr
 - p_1 p_d &  - p_2 p_d &\dots & p_d(1-p_d)}\right].
$$
To obtain the average mean
quadratic error matrix, we integrate $W$ on the rotation invariant states $H$ with
respect to the normalized Haar measure $\mu$. We have to calculate two types of integrals:
$$
\int_H  - p_i p_j  d\mu(\theta)=-\int_H  \Big(e_i + e_i\< f_i,
\theta\>\Big)\Big(e_j +e_j \< f_j, \theta\>\Big) \, d\mu(\theta)
$$
$$
=-\int_H e_i e_j d\mu(\theta)-e_i e_j\int_H  \< f_i, \theta\>d\mu(\theta)-\int_H e_i e_j\<
f_j, \theta\> d\mu(\theta)-e_i e_j\int_H \< f_i, \theta\>\< f_j, \theta\> d\mu(\theta)
$$
$$
=-e_i e_j+0+0-e_i e_j  \alpha \< f_i, f_j\>.
$$
The integrals in the middle are zeros because of the symmetry: $\int_H \< v,\theta\>\,
d\mu(\theta)=0$. The last integral comes from:
$$
\int_H \< f_i, \theta\>\< f_j, \theta\>d\mu(\theta)=\int_H
\left(\sum\limits_{\ell=1}^d f_{i,\ell} \theta_\ell\right)
\left(\sum\limits_{m=1}^d f_{j,m} \theta_m\right)
d\mu(\theta)=\sum\limits_{\ell=1}^d \left(\int_H \theta_\ell^2 d\mu(\theta)\right) f_{i,\ell}
f_{j,\ell}
$$
The quantity $\int_H \theta_\ell^2 d\mu(\theta)$ does not depend on $\ell$: it is a constant 
$\alpha$ depending on the domain of the integration; this way we obtained the stated formula.

Similarly,
$$
\int_H  p_i (1-p_i)  d\mu(\theta)=e_i(1-e_i)-e_i^2 \alpha \< f_i, f_i\>,
$$

so we can calculate the average quadratic error matrix:
$$
F^{-1}
\left[\begin{array}{cccc}e_1^{-1}-1-\alpha  \< f_1, f_1\> & -1-
\alpha  \< f_1, f_2\> & \cdots & -1- \alpha \< f_1, f_d\>\\
 -1 -\alpha \< f_1, f_2\> & e_2^{-1}-1 - \alpha \< f_2, f_2\> & \cdots &
-1- \alpha \< f_2, f_d\>\\
 \vdots & \vdots & &\vdots \\
 -1- \alpha \< f_1, f_d\> & -1- \alpha \< f_2, f_d\> &
\cdots  & e_d^{-1}-1 -\alpha  \< f_d, f_d\>  \end{array}\right]
 (F^{-1})^*,
$$
where $F=(f_1,f_2,\dots, f_{d})$.

The minimizer should be symmetric, hence $e_i=1/{n^2}$,
$\<f_i,f_i\>=x$ and $\<f_i,f_j\>=y$, if $i\ne j$, $i,j \le n^2-1$.
Then the determinant has the form $A/B$, where
$$
A=(n^2-\alpha(x-y))^{n^2-2}(1-\alpha(x+(n^2-2)y))
$$
and
$$
B=(x-y)^{n^2-2} \cdot (x+(n^2-2) y)
$$
Therefore, we minimize
\begin{equation}\label{AperB}
\frac{A}{B}=\left(\frac{n^2}{x-y}-\alpha\right)^{n^2-2}\left(\frac{1}{x+(n^2-2)
y}-\alpha\right).
\end{equation}

We can also calculate the length of
$f_{n^2}=-\sum\limits_{i=1}^{n^2-1} f_i$:
\begin{equation}\label{ineq1}
\<f_{n^2},f_{n^2}\>=(n^2-1) \cdot (x+(n^2-2) y) \le n^2-n
\end{equation}
where the latter inequality is the condition for positivity, see Lemma \ref{L:poz}.
On the other hand, we have the condition
\begin{equation}\label{ineq2}
x \le n^2-n.
\end{equation}
If both inequalities were sharp, then (\ref{AperB}) would not be
minimal, because we could increase both $x$ and $y$ with a
sufficiently small $\varepsilon$ and then the value of
(\ref{AperB}) would be smaller. If the equality holds in
(\ref{ineq1}), then the second term of (\ref{AperB}) will be
constant. So we want to have the difference of $x$ and $y$ as large as
possible, hence $x$ should be maximal (and then $y$ is minimal). If
the equality holds in (\ref{ineq2}), then from (\ref{ineq1}) we
have $y\le -(n^2-n)/(n^2-1)$, and on this domain (\ref{AperB}) has a
negative derivative. So the minima are taken at
$$
x=n^2-n, \quad \textrm{and} \quad y=-\frac{n^2-n}{n^2-1}.
$$
Lemma  \ref{L:poz} gives
$$
E_i=\frac{1}{n^2}(I + f_i\cdot \sigma)=\frac{1}{n}P_i
$$
with some projections $P_i$ and
$$
\Tr P_iP_j=\frac{1}{n^2}\<I+f_i\sigma,I+f_j\sigma\>=\frac{1}{n}+\frac{1}{n^2} y=\frac{1}{n+1}.
$$ We arrived at a
system  (\ref{E:SIC}). The following statement is obtained.

\begin{theorem}
If a  symmetric informationally complete system  exists, then the optimal POVM
is described by its projections $P_i$ as $E_i=P_i/n$ ($1 \le i \le n^2$).
\end{theorem}

 For a qubit the existence of the symmetric informationally complete POVM is obvious,
there are other known examples in lower dimensions, but in $n$ dimensions the existence 
is not known.

\section{A non-symmetrical qubit case for POVM}

In contrast to the previous section where we constructed an optimal
POVM without any information on the state, in the upcoming part we
will show some special, non-trivial cases, where some information
is known. This is a much harder task, therefore the achieved
results are not so strong, nevertheless, it is even more
interesting than the fully unknown case.

Let us assume, first of all, that we have a qubit: $\theta_1,
\,\theta_2$ are not known, but $\theta_3$ is given, so the density operator is only 
partially unknown. In this case the POVM will have three components $\{E_1,E_2,E_3\}$, the
sufficient number of components to make the state estimation. Of
course we can use a greater number of components but in that case
the construction of the estimator is not trivial.

We use the parametrization
$$
E_1=a_0 (I+ a \cdot \sigma), \quad  E_2=b_0 (I+ b \cdot \sigma),
\quad  E_3=c_0 (I+ c \cdot \sigma),
$$
where $a_0,b_0,c_0 \in \bbbr$, and $a,b,c$ are vectors in $\bbbr^3$. From $E_1+E_2+ E_3=I$, we have
$ a_0 + b_0 + c_0=1$ and $ a_0 a+ b_0b + c_0c=0$.
The positivity conditions are:
$$
0 \le a_0, b_0, c_0 \qquad \|a\|, \|b\|, \|c\| \le 1.
$$

Similarly to the previous section we should minimize a determinant of the form $A/B$,
where
$$
A= \Det V, \quad B=\Det^2 \left[ \begin{array}{ccc} a_1& a_2 \\
b_1& b_2\end{array} \right]=\Det
\left[\matrix{ \<a^*,a^*\> & \<a^*,b^*\> \cr \<a^*,b^*\> & \<b^*,b^*\>}\right]=:\Det C,
$$
$$
V=\left[\matrix{ pa_0^{-1}-p^2 & - pq \cr -pq  & q b_0^{-1}-q^2}\right]
-\beta (1-\theta_3^2)\left[\matrix{ \<a^*,a^*\> & \<a^*,b^*\> \cr 
\<a^*,b^*\> & \<b^*,b^*\>}\right]=:D- c\cdot  C
$$
$$
p=1+a_3 \theta_3, \quad q=1+b_3 \theta_3, \quad a^*=(a_1,a_2),\quad b^*=(b_1,b_2),
$$
where $\beta$ is a constant depending on the particular unitarily invariant surface of 
the integration, $0\le \beta \le 1$, so $c$ is a given constant on interval $[0,1-\theta_3^2]$. 

For the minimizer we can assume some symmetry conditions: $a_0=b_0=c_0=\frac{1}{3}$, and
we will minimize $A$ and maximize $B$ independently to obtain the optimum: 
\begin{equation}\label{minA}
A=(d_{11}-c~ c_{11})(d_{22}-c~ c_{22})-(d_{12}-c~ c_{12})^2 \rightarrow\textrm{min},
\end{equation}
\begin{equation}\label{maxB}
 B=c_{11} c_{22}-(c_{12})^2 \rightarrow \textrm{max.}
\end{equation}
Let us suppose first that $a_3$ and $b_3$ are given (then the elements of $D$ are constants) 
and we want to optimize the other variables.

We know that
\begin{equation}\label{bound_c}
\<a+ b,a +  b\> =c_{11}+ c_{22}+2 c_{12}+ (a_3+b_3)^2 \le 1.
\end{equation}

If $c_{12}\ge 0$, then from (\ref{bound_c}) we have $c_{11}+c_{22}\le 1$, hence $B\le 1/4$.

If $c_{12} < 0$, then $B$ is maximal if $c_{12}$ is maximal, from (\ref{bound_c}) we have 
an upper bound:
\begin{equation}\label{bound_c12}
c_{12} \le \frac{1- (a_3+b_3)^2-c_{11}- c_{22}}{2}.
\end{equation}

Substituting this upper bound in (\ref{maxB}) we have to maximize it in $c_{11}$ and $c_{22}$. 
Using derivation we can obtain that it is maximal if $c_{11}$ and $c_{22}$ are maximal:
\begin{equation}\label{bound_c1122}
c_{11}=a_1^2+a_2^2 \le 1-a_3^2 \quad \textrm{and} \quad c_{22}=b_1^2+b_2^2 \le 1-b_3^2.
\end{equation}
Substituting this upper bound  in (\ref{maxB}) we get
\begin{equation}\label{abB}
B=\frac{3}{4}-a_3^2 - a_3 b_3 - b_3^2\, .
\end{equation}
which is optimal if $a_3=b_3=0$. Then $B=3/4$, so it is a global optimum.

\begin{lemma}
The following inequality is always true: $$d_{12}-c~ c_{12}\le 0$$
\end{lemma}

\proof
We have
$$
d_{12}-c~ c_{12}=-(1+\theta_3 a_3)(1+\theta_3 b_3)-c (a_1 b_1+a_2 b_2).
$$
Since $c\le 1-\theta_3^2$ and
$$
a_1 b_1+a_2 b_2 =\<a^*,b^*\> \ge  -\|a^*\| \|b^*\| \ge -\sqrt{1-a_3^2} \sqrt{1-b_3^2},
$$
so it is enough to show that
$$
(1+\theta_3 a_3)(1+\theta_3 b_3) \ge (1-\theta_3^2)\sqrt{1-a_3^2} \sqrt{1-b_3^2}.
$$

The right-hand side does not depend on the signs, the left-hand side is minimal if 
$\theta_3 a_3 \le 0, ~\theta_3 b_3 \le 0$, so it suffices to prove for positive 
$a_3, b_3, \theta_3$ that
$$
(1-\theta_3 a_3)(1-\theta_3 b_3) \ge (1-\theta_3^2)\sqrt{1-a_3^2} \sqrt{1-b_3^2}.
$$
This is true since from the Cauchy-Schwarz inequality follows that
$$
a_3 \theta_3+\sqrt{1-a_3^2}\sqrt{1-\theta_3^2}\le 1 \quad \Longrightarrow \quad 
1-a_3 \theta_3 \ge \sqrt{1-a_3^2}\sqrt{1-\theta_3^2}, 
$$
and a similar statement is true for $b_3$. \qed

Using this lemma, we get that $A$ is minimal if $c_{12}$ is maximal, and from there 
the solution is almost the same as in the previous case, we only have more complicated 
calculations. We substitute the upper bound (\ref{bound_c12}) in to (\ref{minA}) and 
we can obtain that it is minimal if $c_{11}$ and $c_{22}$ are maximal. Using the bounds 
(\ref{bound_c1122}) we get for $A$ a function of $a_3$ and $b_3$; using differentiation 
we can obtain that $a_3=b_3=0$ gives the optimal solution here, too.

So in both cases we have equality in (\ref{bound_c1122}), so $\<a^*,a^*\> = \<b^*,b^*\>=1$
and equality in (\ref{bound_c12}), so $\<a^*,b^*\> =-1/2$. Since $\|a\|=\|b\|=1$,
$$
\frac{3}{2}E_1=\frac{1}{2}(I+a \cdot \sigma), \qquad \frac{3}{2}E_2=\frac{1}{2}(I+b \cdot \sigma)
$$
are projections. We have $\|c\|=\|a+b\|=1$ and this implies that $3E_3/2$ is a projection
as well.

\begin{theorem}\label{T:2.5}
The optimal POVM for the unknown $\theta_1$ and $\theta_2$ can be
described by projections $P_i$ $(i=1,2,3)$:
$$
E_i=\frac{2}{3}P_i, \quad \sum_{i=1}^3 P_i=\frac{3}{2}I,
\quad \Tr \sigma_3 P_i=0, \quad \Tr P_iP_j= \frac{1}{4}
\quad \mbox{for}\quad  i \ne j.
$$
\end{theorem}

The optimal POVM is quasi-orthogonal to the
subalgebra generated by $\sigma_3$ and symmetric in the other
directions. This result is in some sense the combination of Theorem 1 and Theorem 2,
with complementarity and symmetricalness both occurring in the same
problem. The conditions of (\ref{E:SIC}) are valid, but with
different constants than in the SIC-POVM case. 

It is an interesting question whether the existence of a conditional SIC-POVM in 
Theorem \ref{T:2.5} is just a coincidence and it follows from the geometrical structure 
of $M_2(\bbbc)$, or there are such objects in higher dimensions. The following example 
solves this problem:

\begin{example}
Let the elements of a POVM in $M_3(\bbbc)$ be
$$
E_1=\frac17 \left[\matrix{ 1&1&1\cr 1&1&1\cr 1&1&1}\right],
\quad
E_2=\frac17 \left[\matrix{1&\ep^6&\ep^2\cr \ep&1&\ep^3\cr \ep^5&\ep^4&1}\right],
\quad
E_3=\frac17 \left[\matrix{ 1&\ep^2&\ep^3\cr \ep^5&1&\ep\cr \ep^4&\ep^6&1}\right],
$$
$$
E_4=\frac17 \left[ \matrix{1&\ep^4&\ep^6\cr \ep^3&1&\ep^2\cr \ep&\ep^5&1}\right],
\quad
E_5=\overline{E_2}, \quad E_6=\overline{E_3},\quad E_7=\overline{E_4}, 
$$
where  $\ep=\exp\left( 2\pi \im / 7 \right)$.
This POVM fulfills the conditions in (\ref{E:SIC}) with constants $k=7$, $\lambda=7/3$ and 
$\mu=2/9$. Moreover it is quasi-orthogonal to diagonal matrices, so it is a conditional 
SIC-POVM with respect to the diagonal matrices. \qed
\end{example}

It is still an open question, however, that exactly in which cases a conditional SIC-POVM 
exists. Furthermore, does it minimize the determinant of the average covariance matrix? The 
answer to the first question is definitely a difficult problem, because it contains the 
question of the existence of SIC-POVMs. Our conjecture to the second question is that 
conditional SIC-POVMs are optimal if they exist, and so they can play an important role 
in quantum information theory, but an exhaustive investigation exceeds the scope of this paper.

\section{Discussion and conclusions}

A density matrix $\rho \in \Mn$ has $n^2-1$ real parameters. If
there are no known parameters, several two-valued POVMs
$\{F_1,I-F_1\},\, \{F_2,I-F_2\},\, \dots
\,,\{F_{n^2-1},I-F_{n^2-1}\} $ or a single POVM $\{E_1,E_2,\dots,
E_{n^2}\}$ are considered. The estimator can be constructed as a
linear function of the outcomes, and since they are multinomially
and binomially distributed, their variance can be calculated
easily. By averaging them on the invariant states and taking the
determinant of the average covariant matrix, we will get
comparable results. The optimum is at the quasi-orthogonal
measurements (Theorem 1) and at a symmetric informationally
complete system (Theorem 2), respectively. If partial information
is known, the concept is the same as previously, but as there are
some known parameters, the domain of integration is restricted.
From Theorem 1, it turns out that in the von Neumann case the
optimal measurements are not only complementary to each other, but
to the known subalgebras as well. For POVMs a qubit case is
described in detail: conditionally symmetric informationally complete systems turn
out to be optimal. The latter is a new concept, an example for the qutrit case is obtained
too, further investigations are suggested.

\end{document}